\documentclass[10pt,twocolumn]{IEEEtran}

\usepackage{graphicx}
\usepackage{amssymb}
\usepackage{amsfonts}
\usepackage[cmex10]{amsmath}
\usepackage{psfrag}
\usepackage{algorithm}
\usepackage{algorithmic}

\usepackage{cite}
\usepackage{listings}
\usepackage{color} 
\usepackage[normalem]{ulem} 
\usepackage{amsfonts,amsmath,amssymb,cite,graphicx,algorithm,delarray}
\usepackage[framed,numbered,autolinebreaks,useliterate]{mcode}
\usepackage{subfigure}

\begin{document}

\title{
\vspace{0cm} Developing and Assessing MATLAB Exercises for Active Concept Learning \thanks{This work was partially supported by the HKUST Teaching Development Fund under Grant No. 080E-1516.} \thanks{ S.H. Song, Marco Antonelli and Bertram E. Shi are with the Electronic and Computer Engineering Department at HKUST. Tony Fung and Albert Lo are with the Hong Kong Virtual University at HKUST, Brandon D. Armstrong is with MathWorks, and Amy Chong is with the Center for Education Innovation at HKUST.}}
\author{S.H. Song, Marco Antonelli, Tony Fung, Brandon D. Armstrong, Amy Chong, Albert Lo, Bertram E. Shi}
\maketitle

\begin{abstract}
New technologies, such as MOOCs, provide innovative methods to tackle new challenges in teaching and learning, such as globalization and changing contemporary culture and to remove the limits of conventional classrooms. However, they also bring challenges in course delivery and assessment, due to factors such as less direct student-instructor interaction. These challenges are especially severe in engineering education, which relies heavily on experiential learning, such as computer simulations and laboratory exercises, to assist students in understanding concepts. As a result, effective design of experiential learning components is extremely critical for engineering MOOCs. In this paper, we will share our experience gained through developing and offering a MOOC on communication systems, with special focus on the development and assessment of MATLAB exercises for active concept learning. Our approach introduced students to concepts using learning components commonly provided by many MOOC platforms (e.g., online lectures and quizzes), and augmented the student experience with MATLAB based computer simulations and exercises to enable more concrete and detailed understanding of the material.  We describe here a systematic approach to MATLAB problem design and assessment, based on our experience with the MATLAB server provided by MathWorks and integrated with the edX MOOC platform. We discuss the effectiveness of the instructional methods as evaluated through students’ learning performance. We analyze the impact of the course design tools from both the instructor and the student perspective.
\end{abstract}


\section{Introduction}

Many forces are driving the need to update engineering education: globalization, the availability of new technologies, changing contemporary culture, etc. New technologies, such as massive online open courses (MOOCs), present both opportunities and challenges for engineering education. On one hand, online pedagogic tools remove many limits of conventional classrooms, such as the need for synchronization (time) and co-location (space), limitations on class size, and the one-size-fits-all problem \cite{NAE}. The role of e-learning and students' motivation to use e-learning/online education have been discussed in \cite{Tot, Thomas}.  On the other hand, online tools have disadvantages, such as less direct student-instructor interaction and difficulty in the design of effective course assessment, which lead to issues like high drop-out rate \cite{Thanasis}. By creating new learning experiences, MOOCs are driving new research into pedagogical innovation in course design and assessment to maximize the effectiveness of online learning\cite{Dragana}\cite{Dilrukshi}.

MOOCS present particularly serious challenges to engineering courses \cite{Sajid}, which rely upon experiential learning through hands-on experiments and computer simulations.  Simulations are a very important component of the modern engineering curriculum. De Jong and Van Joolingen have related simulations to scientific discovery learning where students learn by bridging the process of discovery and the process of learning \cite{DeVan}.  Computer simulation is an extremely successful example of how technology improves research, teaching, and learning.  From the research point of view, it is a powerful tool to validate theories and observe/explore phenomena not easily accessible due to technical or economic constraints.  From the teaching point of view, it can be a way to ground and provide insight into abstract concepts and to develop skills, such as modeling \cite{HEER}. From the learning point of view, it allows students to discover the effect of changing a parameter or some aspect of an algorithm quickly and without the need for a real (and sometimes inaccessible) system. It also enables students to learn the art of modeling a complex system, which is a fundamental skill in the engineering field \cite{Kruger_2011}.

The difficulty of laboratory setup \cite{Chenard}, running remote laboratories \cite{Gabriel, Felix} and computer simulation components in MOOCs have been well recognized \cite{Sajid}. Assessment of laboratory exercises also presents challenges.  Assessment normally serves three purposes \cite{HEER}: assisting student learning, measuring student understanding and mastery, and evaluating the effectiveness of university programs. Course and assessment design for assisting learning relies heavily on theories and models of student learning, as well as statistical measurement and inference. This paper is primarily concerned with the first purpose: assisting student learning.

This paper focuses on the design of interactive computer simulation experiments in MATLAB, and the automated assessment of student work in these experiments. It is based upon our experience offering the MOOC, HKUSTx ELEC1200: A System View of Communication from Signals to Packets, on edx.org. This MOOC is based upon an introductory course to the curriculum offered by the Department of Electronic and Computer Engineering (ECE) at the Hong Kong University of Science and Technology (HKUST).  These courses use the context of wireless communications to introduce students to important ECE concepts, such as linear systems theory, probability, signal processing and networking.  In the laboratory exercises in the physical course, students use MATLAB to generate waveforms to be transmitted over an infrared communication channel, and to process the received waveforms. These laboratories play an important role in students' learning by allowing students to understand abstract concepts through building and testing a communication system. In the MOOC version of this course, we have replaced the actual physical infrared communication channel with a computer simulated channel. However, the key learning components of the laboratories: writing MATLAB code to generate and process waveforms are largely the same.  Although efficient teaching and learning methods for communication courses \cite{Wei, Thomas1, Yair} have been studied previously, they focused on specific platform (LabVIEW) \cite{Wei}, simulation exercises \cite{Thomas1}, or (FPGA based) simulation board \cite{Yair}. In this work, we describe systematic methodology to design simulation task and assessment code for communications courses, and analyze the impact, opportunities, and challenge of different design platforms.

We describe our approach to address a number of challenges.

First, computer simulation based lab exercises may have multiple complementary goals. In developing the MATLAB exercises for our course, we had two primary objectives: (1) we wanted to improve students' understanding of the courses concepts through experiential learning and (2) we wanted to teach students how to use MATLAB for simulation.

Second, there are technological challenges as well. Since the MOOC platform we used to offer this course (edx.org) and all other MOOC platforms that we are aware of cannot execute MATLAB code, the development of these exercises must be done by integrating the information distribution capability of the MOOC platform with the capability to execute MATLAB code of servers managed at MathWorks.  This requires that different parts of the laboratory exercises be executed on different servers, while at the same time presenting students with a unified learning experience. Here we describe two different design platforms for developing such exercises.

Finally, the assessment of computer simulations in MOOCs is challenging because, different from on-campus labs, feedback and assessment are automatically performed by remote servers.  The automated assessment of computer simulations is not trivial \cite{Johnson}. Unlike the quiz questions commonly used to aid and assess understanding, for computer simulations:
\begin{itemize}
\item The answer is not a number or an English character as for the MC (multiple-choice) or fill-in-the-blank cases.
\item The answer is not unique, because there are different ways to implement the same algorithm.
\end{itemize}
At a minimum, the automated grader should decide whether the submitted solution executes without syntax or programming error and produces the correct result. Ideally, it should also be able to identify the most common mistakes \cite{Hestenes} provide feedback messages that can guide students to correct errors in their submission without indicating the exact error in their submission. Despite some progress in automatic grading for computer programming/simulations, more effort is needed \cite{Shashank}. Since our primary goal was to assist understanding of the course concepts, our assessments focused primarily on checking the logical correctness of the program, and not on other aspects of programming, such as style and efficiency.  Nonetheless, students did learn about these other aspects through issues raised in the discussion section of the MOOC.

The specific contributions of this paper are:
\begin{itemize}
\item We present a systematic framework for design of computer simulation based lab exercises as a number of discrete tasks.
\item We describe methodology for designing each task.
\item We propose a systematic assessment code design methodology.
\item We evaluate the impact of the two design platforms we used from both the instructor and the students' perspectives.
\item We analyze the opportunities and challenges of new communication tools for pedagogic use.
\end{itemize}

\section{Course Description}

HKUSTx ELEC1200: A System View of Communications from Signals to Packets, is an introductory course to Electronic and Computer Engineering, which covers core concepts from the systems side of the ECE curriculum, and motivates their importance by illustrating their use in the design and optimization of a wireless communication system.  The course is divided into three parts. Part I covers a simple point to point link between a transmitter and receiver over a discrete time channel.  It first identifies several channel effect, including transmission delay, channel distortion, and additive noise.  It then introduces ways to describe, understand and ultimately handle these issues, including communication protocols, linear time invariant channel models, equalization, and channel coding. Part II focuses on sharing a channel using frequency division multiplexing and both analog and digital modulation.  This provides the opportunity to introduce frequency domain analysis of signals. We revisit the concept of equalization, previously introduced using a purely time domain perspective. We also describe the concept of source coding. Part III focuses on communication networks, and introduces the layered network structure. Given the extensive coverage of the physical layer in Parts I and II, Part III focuses on the the link, network, transport, and application layers. The functions of each layer, e.g. multiple access, routing/forwarding, reliable transport, and support for different applications are introduced.

The course is built around five learning components: lecture videos, quiz questions, lab demo videos, lab exercises, and a final exam. Students earn credits by completing the quizzes, lab exercises, and the final exam, which account for 20, 30, and 50 percent of the final grade, respectively. To qualify for a verified certificate, students cumulative score must exceed 60$\%$.

In each week, students first watch a sequence of short lecture videos covering important communications concepts. These videos are typically about 10 minutes in length, and cover one topic.

\begin{figure}
    \begin{center}
    \scalebox{0.45}{\includegraphics{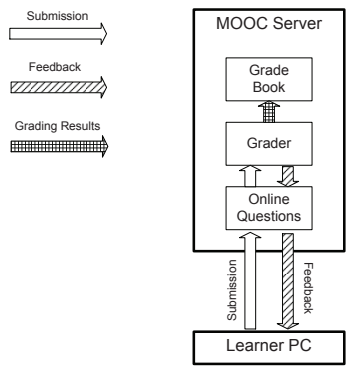}}
    \end{center}
    \caption{Data flow for quiz questions.}
    \label{Quiz}
\end{figure}

After each lecture video, there are quizzes containing MC questions, and/or fill-in-the-blank questions. These serve to check students' understanding of the concepts covered in the videos. Quiz grading is done automatically by the MOOC platform. Most MOOCs platforms provide simple graders for quiz questions, such as expression graders \cite{Johnson}. The network structure and data flow for grading quiz questions is shown in Fig. \ref{Quiz}. Students read the questions from the MOOC web page and fill in their answers directly on the page. The MOOCs server will then grade the solutions, give feedback to students, and record the grading results.

Students are also presented with one or more MATLAB-based lab exercise per week.  Typically, this exercise involves simulating some part of a communication system. For each lab exercise, students first watch a lab demo video summarizing the concepts required to complete the lab exercise, and identifying part of the communication system being simulated within the context of the entire communication chain.  Each lab exercise consists of three to four programming tasks. Student work is assessed automatically, as described below.

The final exam is divided into sections, each covering about one week of material. Each section contains a mixture of MC, fill-in-the-blank, and MATLAB-based questions.

\section{Design of MATLAB-based lab exercises}
\label{SectionDesign}

In this course, there are several intended learning outcomes (ILO) for the MATLAB-based lab exercises. Through these exercises, students will
\begin{enumerate}
\item[ILO1.] Observe how a communication system works and develop an understanding of communications concepts and models.
\item[ILO2.] Develop system design skills through simulation.
\item[ILO3.] Develop problem solving skills, including data interpretation, hypothesis generation and testing, etc.
\item[ILO4.] Develop simulation and programming skills.
\end{enumerate}

In order to achieve these outcomes, we designed the simulations to be detailed enough that the concepts are adequately illustrated from the learner point of view. At the same time, the level of abstraction should be high enough that students are not distracted by unnecessary details.  We want students to spend relatively less time understanding the code, and more time focusing on the learning objective. Balancing these two objectives is critical in the design of a successful set of exercises.

Each week in the MOOC course covered two topics.  Each lab exercise covered the material in about one or two topics. For example, over the course of Part I, which covered a digital communication over a bandlimited and noisy point to point link, the lab exercises were:
\begin{enumerate}
\item A Communication Example
\item Step Response
\item Communication Protocol
\item Performance Evaluation
\item Eye Diagram
\item Equalization
\item Noise and Bit Error Rate
\item Repetition Codes
\item Parity Codes
\end{enumerate}

\subsection{Multi-task decomposition}
\label{multi-task}

In order to achieve the ILO's above, we decomposed each lab exercise into several different types of tasks.  The first task is an overview task to help orient the students.  The next few tasks are implementation tasks, where students must write code implementing one of the functional blocks presented in the overview task.  The final task is typically a performance evaluation task, where students study how the performance of a communication system changes as its parameters change.  We describe each of the task types in more detail below.

The goals of the first overview task are to help students to
\begin{enumerate}
\item Identify the section of the overall communication chain being studied.
\item Understand the overall flow of signals and information between the functional blocks.
\item Understand the input, output and transformation performed by each block.
\end{enumerate}

In this task, students are typically provided with a correctly working MATLAB script simulating the section of the communication chain being studied.  The main components making up this section are encapsulated into functions.  We use function encapsulation so that students can focus on the goals of the task without being distracted by the detailed implementation of the different blocks, which is covered in later tasks.

Students can run the script by pressing a ``Run'' button.  The MATLAB script is executed on a MATLAB server, and the results are sent back to the MOOC server for display.  When students run the code presented in the overview task, the code generates figures showing the signals at different positions of a communication system, e.g. the bit input to the transmitter, the input and output of the channel, and the bit output of the receiver (ILO1). Students are also instructed to change relevant parameters (e.g. the bit time) and observe their effect on these signals (ILO2 and ILO3).

In an implementation task, students are asked to write the code implementing one of the functional blocks presented in the overview task. The goal of the implementation task is to help students to understand exactly how the functional blocks transforms its input to its output. For example, in the lab exercises covering error correcting coding, there are two tasks, where students are asked to implement the encoder in one task and the decoder in the other.

The overall code presented to the students in an implementation task is nearly identical to that in the overview task.  However, the function being implemented is replaced by an incomplete or incorrect code segment.  Importantly, in all tasks, the initial code presented to the students contains no syntax errors and is fully executable.  The output variable of the function to be implemented is defined, albeit incorrectly.  Thus, when the students click on the ``Run'' button, there are no run-time errors generated, and the code still generates figures for display.  However, students can identify that the code is incomplete or incorrect because the contents of those figures differ from that generated in the overview task.

It is the students' job to complete or correct the code implementing the missing function. We make the code in these tasks similar to that in the overview task so that students can compare the contents of the figures generated by the two tasks in order to infer logical errors in their code, and to judge when the code is correct.  Through these tasks, we seek to address ILO3 and ILO4.

Since we do not assume students enrolling in this course are very familiar with MATLAB and coding, we make the exercises progressively more and more difficult. At the beginning of the course, we provide almost the entire code required to implement the function, except for a simple logical error.  The students task is then to find and correct this logical error.  Through reading the code, the students gain insight into MATLAB programming, good programming style, and how to simulate communication systems in MATLAB.  Once students understand the general idea of what the code is trying to do, they can try to infer the error by both thinking through the logic of the function, as well as by comparing the correct and incorrect outputs generated. This allows students to hone in on the error from multiple perspectives. We also provide hints within the written instructions given to the students and links to relevant MATLAB documentation, either maintained on the MOOC site or on the web.  As the course progresses, the errors become more and more complicated to fix and less and less of the code is provided.

The final evaluation task is meant to convey the value of simulation as a way to enhance conceptual understanding and assist learning. It normally requires students to evaluate how the performance of a communication system changes as one of the parameters changes.  For example, they might be asked to use MATLAB to generate a plot of bit error rate as the signal to noise ratio changes.  This gives students insight into various engineering trade-offs (e.g. transmit power versus bit error rate), as well as enabling them to experiment with different techniques to improve performance. Through this final task, we address primarily ILOs 1 through 3.

\subsection{Design philosophy}
\label{philosophy}

In order to best achieve the goals outlined above, we kept the following design philosophy in mind during the design of the lab exercises.

\begin{enumerate}
\item Within each part, all lab exercises are designed around the same basic communication system.  Different lab exercises focus on different concepts, components, or behaviors within this system. This gives students a holistic view, enabling them to see and understand the relationship between different concepts or components.
\item This focus on the same system, enables students to more easily pick up topics with increasing complexity. Since they gain more and more intuition and familiarity with the system, the overhead in understanding the simulations reduces over time.
\item Within each task, the initial provided code should execute as given without syntax or runtime errors. This enables students to focus on the logical flow of each component and the intended learning outcomes, rather than being distracted by the programming.  However, the students are not insulated from syntax or runtime errors, as they often introduce them in the process of completing the code.
\item In the vast majority of cases, we use the code to generate figures, which serve to provide students a concrete visualization of the signals within the communications concepts, and how they change as the architecture or parameters of the system change.  This also serves to help students realize the capacity of using simulations to improve their understanding.
\item Students are required to write code with \emph{increasing} depth and complexity as they progress through the course. For example, in the first task, students use basic MATLAB functions such as vector concatenation to create signals.  In later tasks, they need to use more complex functions like ``\textbf{xor}()'' to implement channel coding.
\item As discussed in more detail below, students are provided with automated grader providing constructive feedback. They are given many chances to submit their solutions for checking and modification. This facilitates a discovery and learning process through trial and error.
\item To provide more flexibility to students, Mathworks made a desktop version of MATLAB available to students by download for use in the coursework during the active class, as well as a library of the higher level functions built for the class. This enabled students to work and experiment on their desktops before submitting their work via the online system.  Since many of the tasks involved implementing the functions provided in the library, we distributed this library as protected MATLAB files (.p files) to prevent students from solving the lab exercises simply by viewing the source code.
\end{enumerate}

\subsection{Illustrative Example}
\label{SecIllustrativeExample}
\begin{figure}
    \begin{center}
    \scalebox{0.35}{\includegraphics{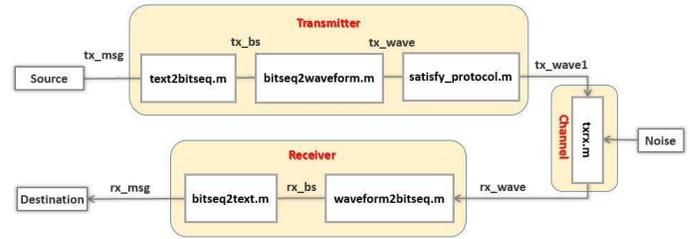}}
    \end{center}
    \caption{System diagram for a communication system.}
    \label{Lab_1_1}
\end{figure}

In the following, we illustrate the design concepts and philosophy outlined previously by describing the first lab exercise of this course, LAB1. This exercise introduces basic communications concepts through work in simulating and implementing a system that communicates a text messages from a transmitter to a receiver over a bandlimited channel. This lab also serves to familiarize students with the online lab environment and with MATLAB.

The text message communication system is shown in Fig. \ref{Lab_1_1}. Students use MATLAB to plot the signals at different points of this communication system, and to implement some of the key functional blocks. We hope that by doing so, students will better understand how these components work together to achieve communication.

In particular, students must complete four tasks:
\begin{itemize}
\item[Task1.] Simulate a simple communication system for sending and receiving a text message.
\item[Task2.] Look into the steps required to encode a text message as a bit sequence and implement the ``Text to Bits'' block (text2bitseq.m).
\item[Task3.] Convert a bit sequence into a discrete-time waveform and implement the ``Bits to Waveforms'' block (bitseq2waveform.m).
\item[Task4.] Convert the received bit sequence to a text message and implement the ``Bits to Text'' block (bitseq2text.m).
\end{itemize}
These are described in more detail below.

\subsubsection{Task 1}

In this overview task, the students are presented with a window pre-populated with the MATLAB code  shown below.

\begin{lstlisting}[basicstyle=\ttfamily\scriptsize]
tx_msg = 'Finished!';   % message to transmit
SPB = 20;               % bit time (samples per bit)
% transmitter %
tx_bs = text2bitseq(tx_msg);          % text=>bit
tx_wave = bitseq2waveform(tx_bs,SPB); % bit=>waveform 
tx_wave1 = satisfy_protocol(tx_wave,SPB); % protocol          
% channel %
rx_wave = txrx(tx_wave1);                 
% receiver %
rx_bs = waveform2bitseq(rx_wave,SPB); % waveform=>bit 
rx_msg = bitseq2text(rx_bs);          % bit=>text 
  
\end{lstlisting}

Each variable corresponds to the signal at some point in the communication system shown in Fig. \ref{Lab_1_1}, and each function corresponds to one block. This enables students to connect the different parts of the code to the graphical representation, so that they can easily get the whole picture. To complete the task, students must complete the following steps.

\begin{figure}
\centering
\subfigure[Transmitted bit sequence and the transmitted wave, SPB=20.]{
    \includegraphics[width=.22\textwidth]{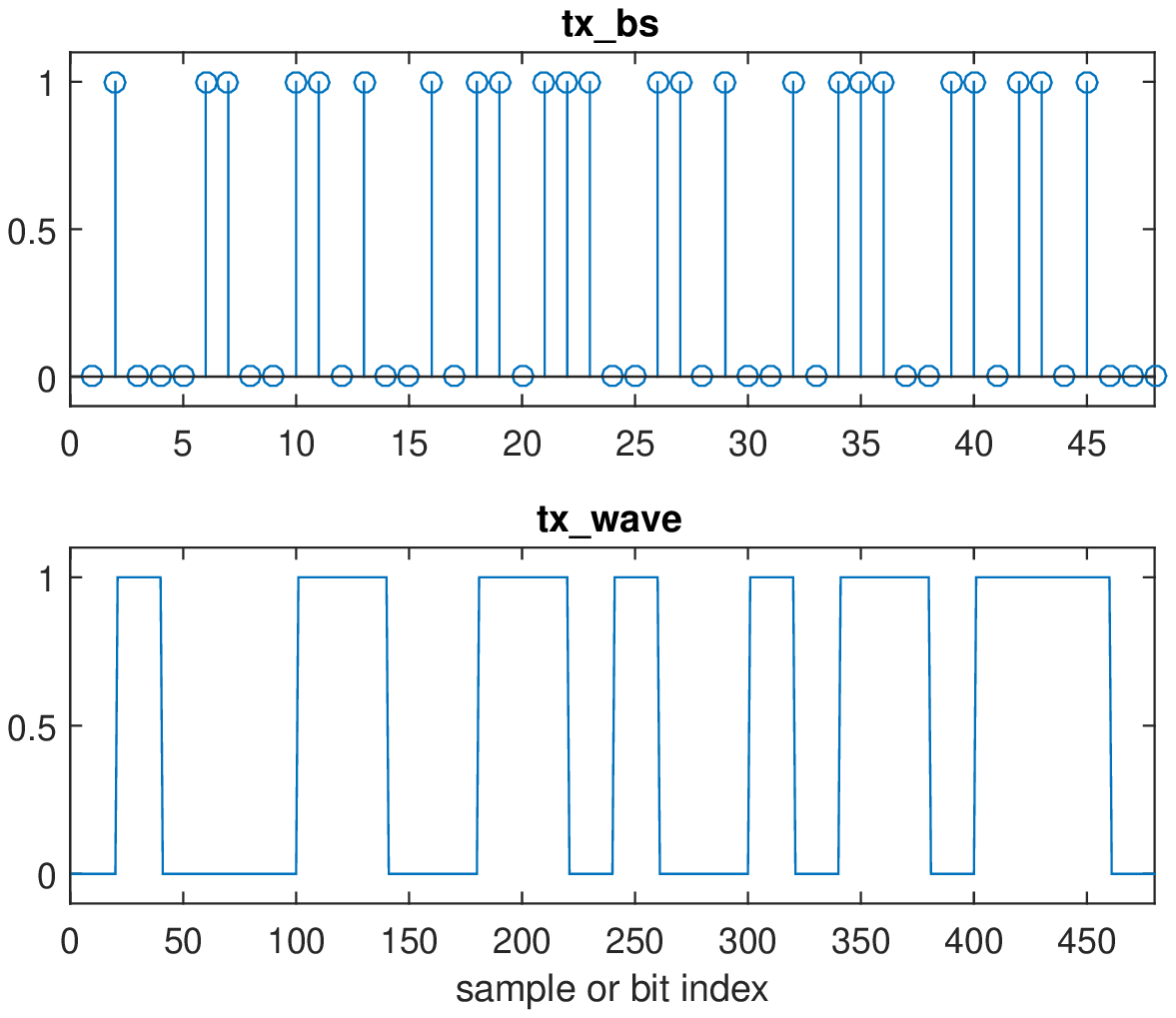}
    \label{s1_tx_bs}
  }
  \subfigure[Received waveform and the received bit sequence, SPB=20.]{
    \includegraphics[width=.22\textwidth]{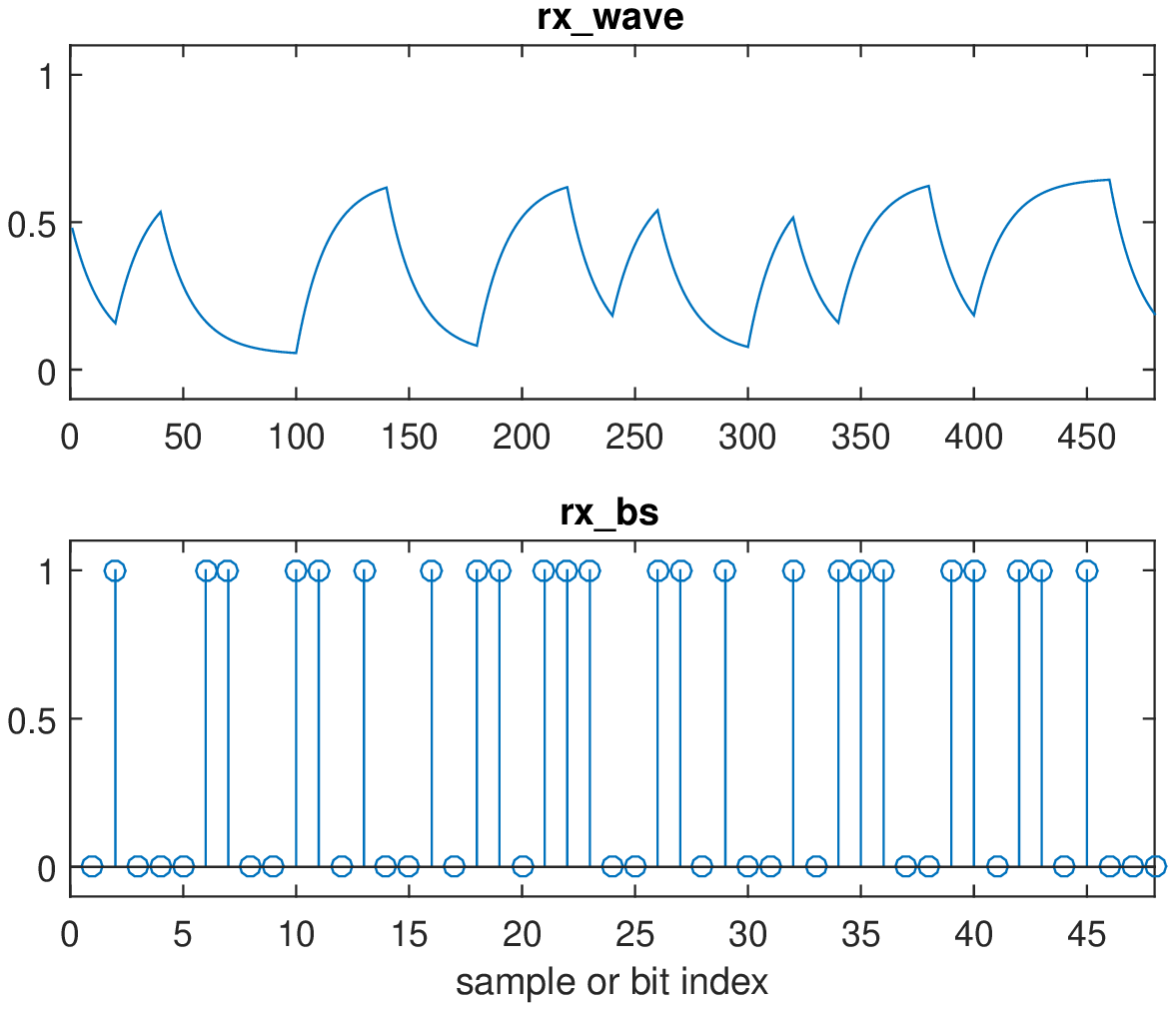}
    \label{s1_rx_wave}
  }
  \subfigure[Transmitted bit sequence and waveform with different msg.]{
    \includegraphics[width=.22\textwidth]{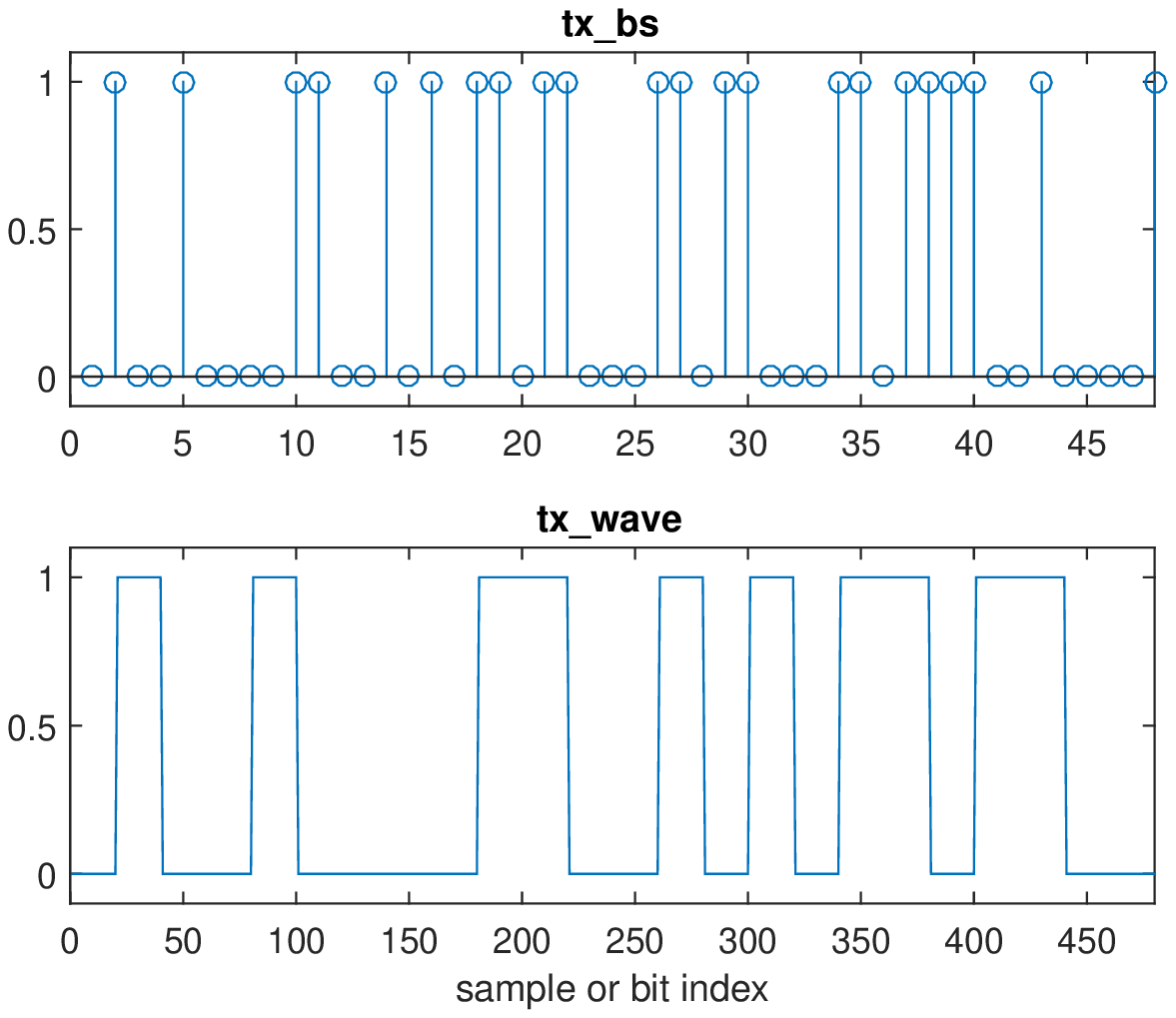}
    \label{s2_tx_bs}
  }
  \subfigure[Received waveform and bit sequence with different msg.]{
    \includegraphics[width=.22\textwidth]{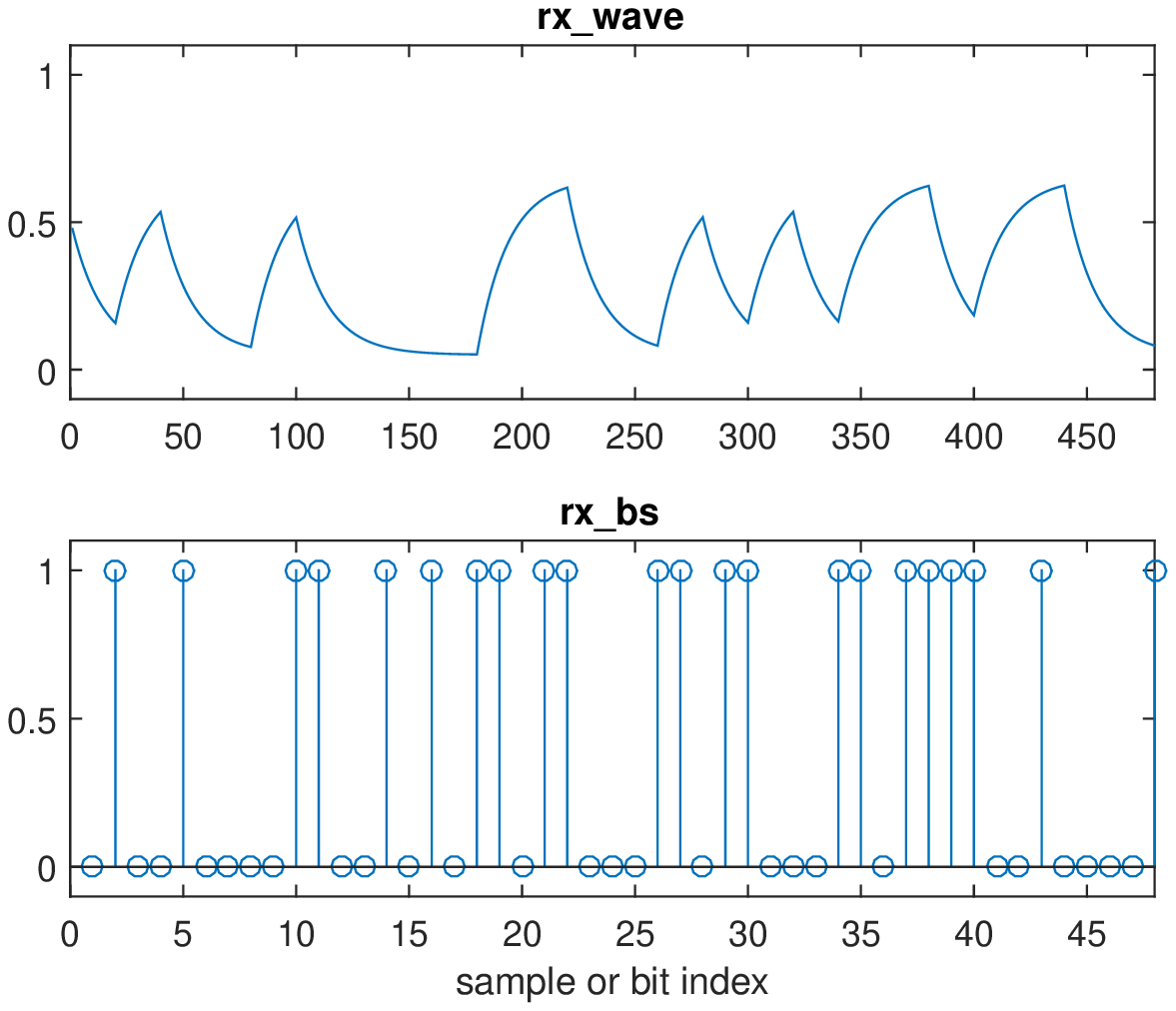}
    \label{s2_rx_wave}
  }
  \subfigure[Transmitted bit sequence and the transmitted wave, SPB=10.]{
    \includegraphics[width=.22\textwidth]{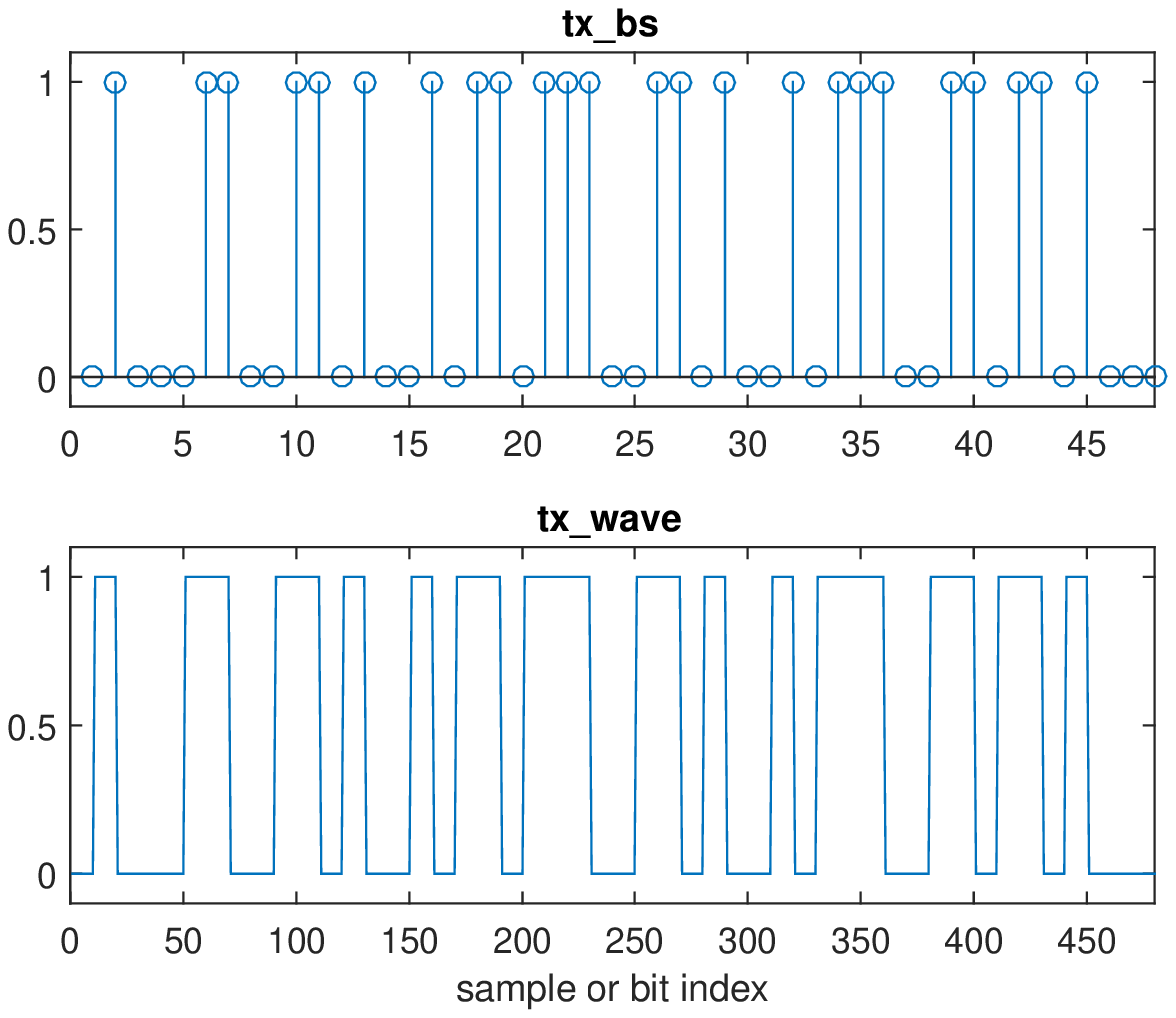}
    \label{s3_tx_bs}
  }
  \subfigure[Received waveform and the received bit sequence, SPB=10.]{
    \includegraphics[width=.22\textwidth]{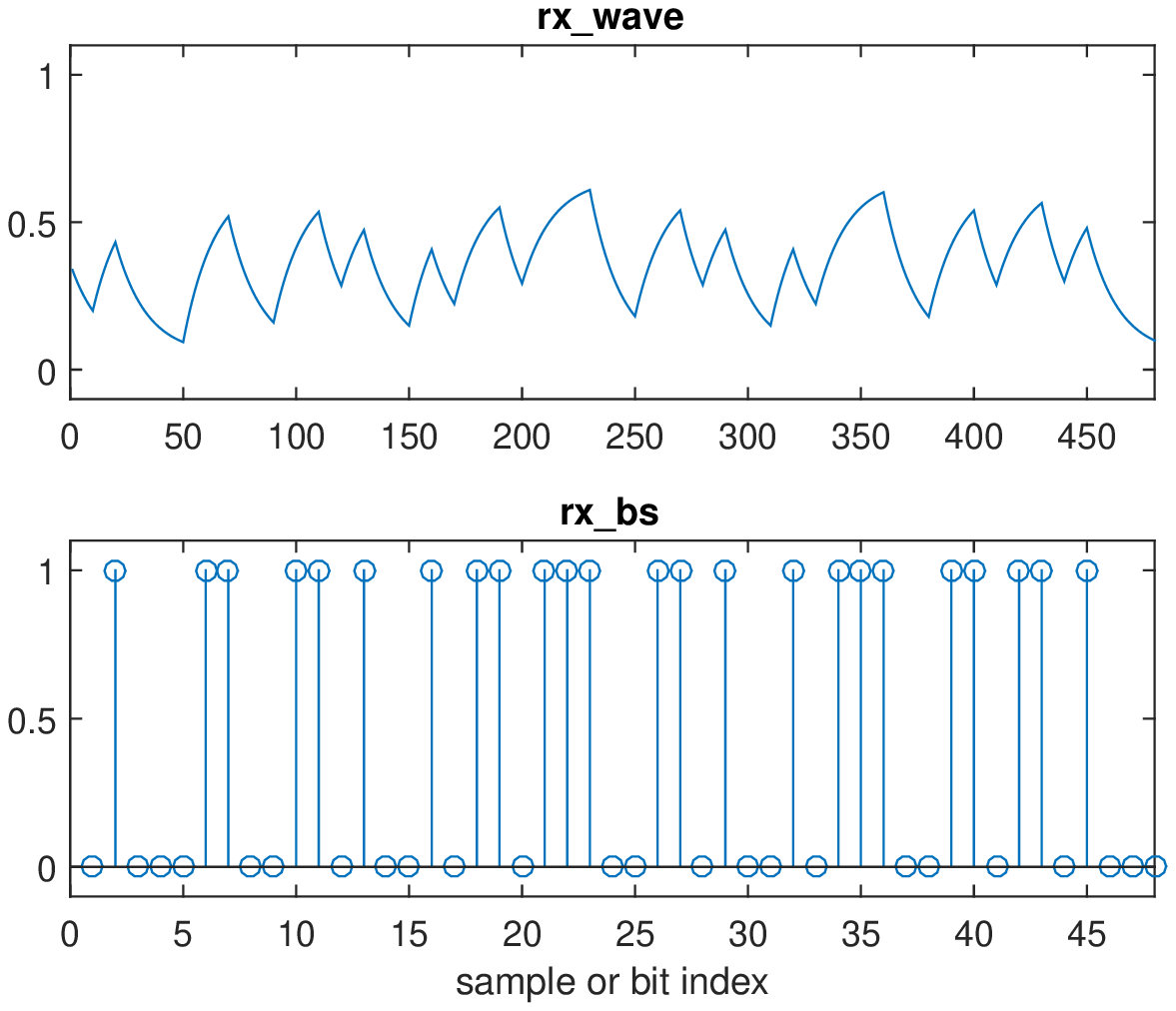}
    \label{s3_rx_wave}
  }
    \subfigure[Transmitted bit sequence and waveform before correction.]{
    \includegraphics[width=.22\textwidth]{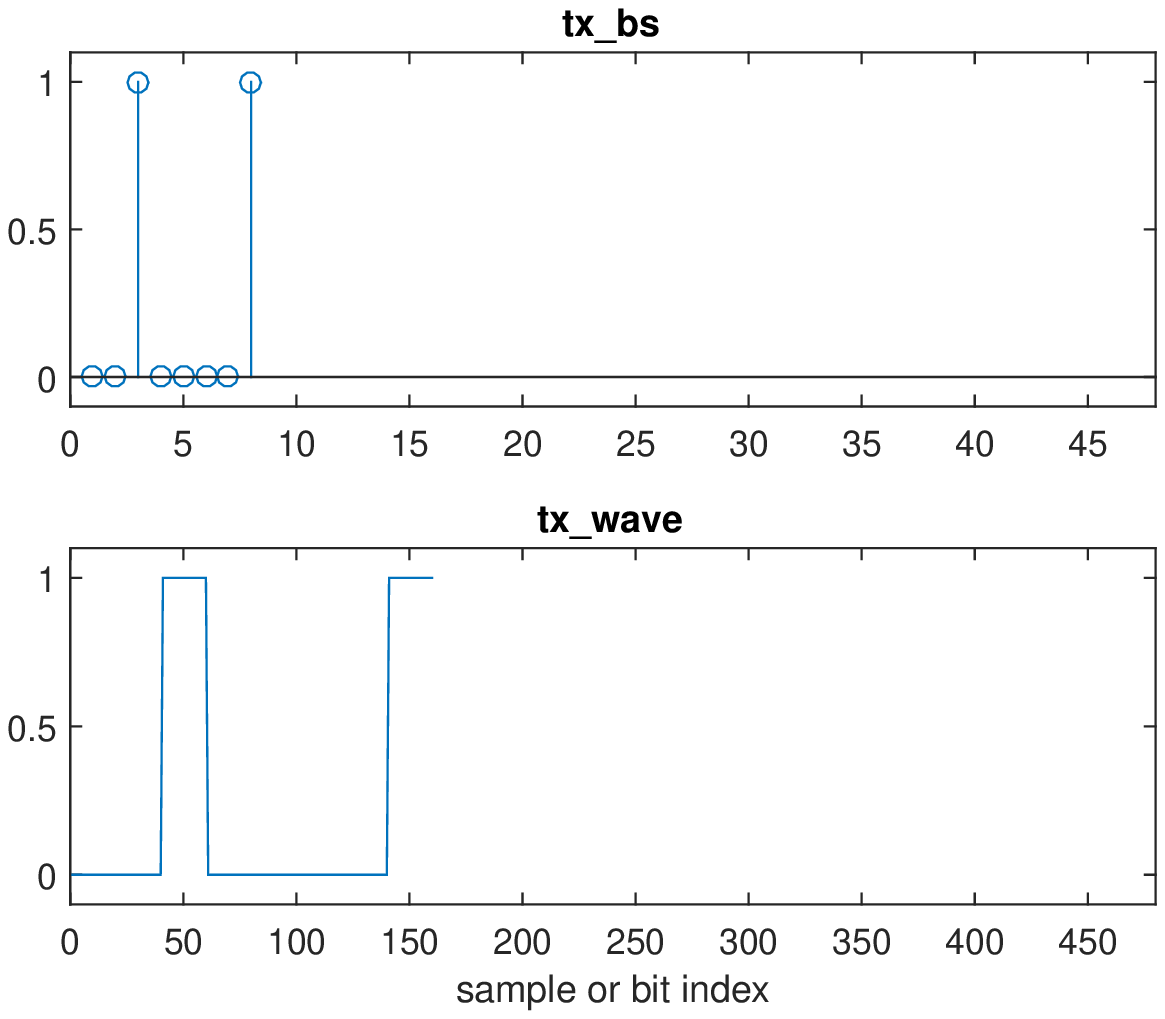}
    \label{t2_tx_bs}
  }
  \subfigure[Received waveform and bit sequence before correction.]{
    \includegraphics[width=.22\textwidth]{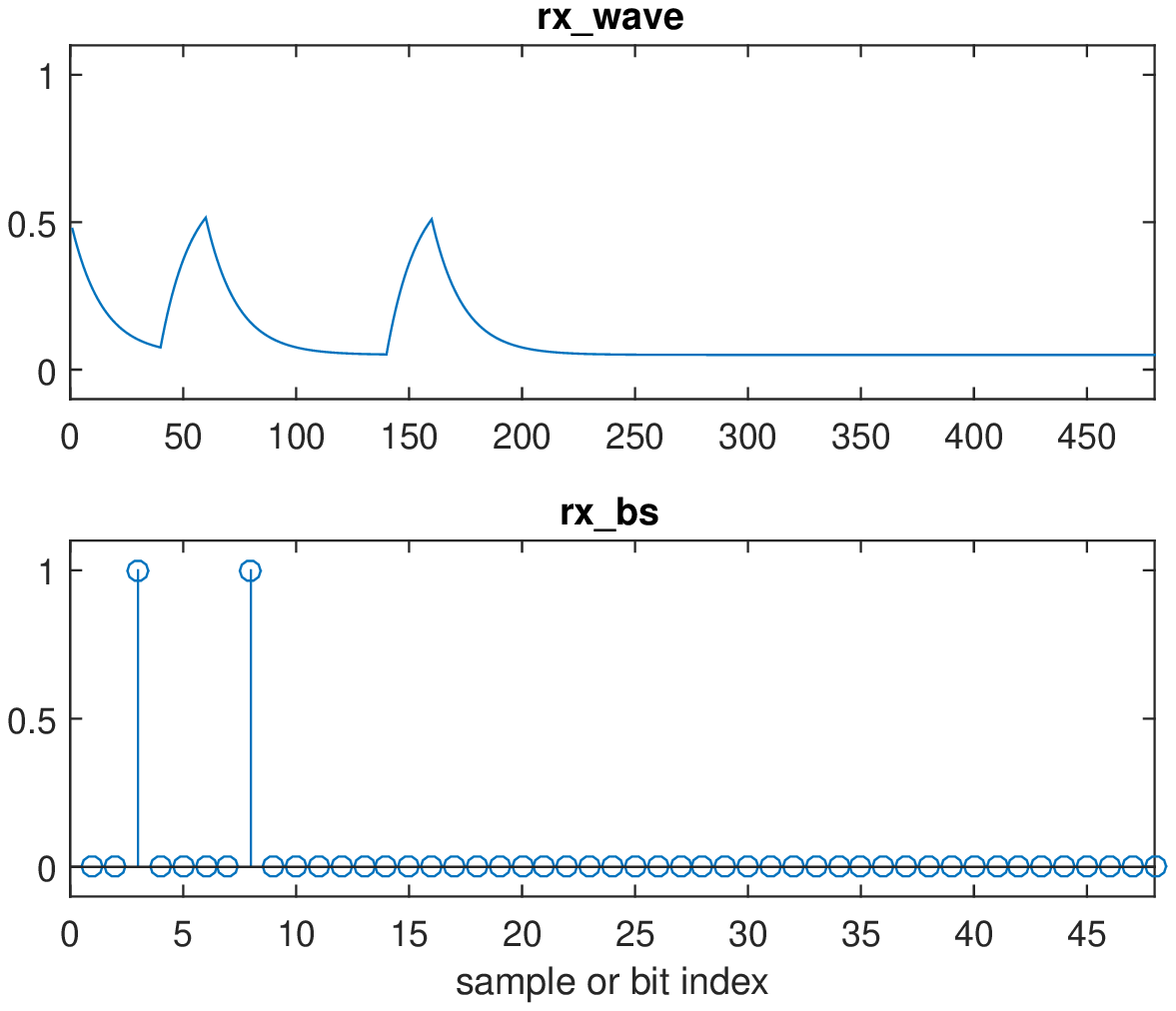}
    \label{t2_rx_wave}
  }
  \caption{Simulation results to show the transmitted and received waveforms.}
\end{figure}

\emph{Step 1: Run the MATLAB code.} We first give a detailed explanation of the code, and then ask students to run the given code by clicking on a ``Run'' button presented below the code window. The MOOC server then submits the code to a MATLAB server that runs the code and returns the outputs generated by the code to be sent back to the MOOC platform for display.

The outputs generated by this code are strings displaying the transmitted and received text messages, and figures plotting the signals at different points of the communication system: the transmitted bit sequence tx$\_$bs, the transmitted wave tx$\_$wave, the received waveform rx$\_$wave, and the received bit sequence rx$\_$bs, as shown in Figs. \ref{s1_tx_bs} and \ref{s1_rx_wave}. By examining these outputs, students get an idea about how information is encoded at different positions of the communication system.

\emph{Step 2: Change the input message.} In this step, students are asked to change the transmitted text message from ``Finished!'' to ``Hello!'' and to observe how this change affects the signals tx$\_$bs, tx$\_$wave, rx$\_$wave, and rx$\_$bs,  shown in Figs. \ref{s2_tx_bs} and \ref{s2_rx_wave}. By comparing the results with those in step 1, students should gain insight into how the message is represented by bit sequences and waveforms by answering questions such as the following. Are the first eight bits of the transmit bit sequence the same? What about the next eight bits? How does the total number of bits in the bit sequence change? How does this affect the length of the transmitted waveform?

\emph{Step 3: Change the bit time.} Students are required to change the bit time (measured in samples per bit) by changing the parameter SPB from 20 to 10 and observe how this change affects the signals at different points of the system. The result is shown in Figs. \ref{s3_tx_bs} and \ref{s3_rx_wave}.

\emph{Step 4: Submit your work.} When students are finished with these experiments, they submit their work for credit.

\subsubsection{Task 2}
In Task 2, students study the implementation of the ``text2bitseq.m'' block shown in Fig. \ref{Lab_1_1}, which takes as input a text string of ASCII characters and produces the corresponding bit sequence.

Students are presented with a window pre-populated with the code shown below.

\begin{lstlisting}[basicstyle=\ttfamily\scriptsize]
tx_msg = 'Finished!';  % message to transmit
SPB = 20;              % bit time in samples per bit

% transmitter %
%------tx_bs=text2bitseq(tx_msg)-------
% % % Revise the following code % % %
tx_bs = [];
for c = 1:length(tx_msg)
    character = tx_msg(c);% get next character from msg
    byte = char2byte(character);% find the 8-bit ASCII
    tx_bs = [byte];
end
% % % Do not change the code below % % %
tx_bs = text2bitseq(tx_msg);          % text=>bit
tx_wave = bitseq2waveform(tx_bs,SPB); % bit=>waveform 
tx_wave1 = satisfy_protocol(tx_wave,SPB); % protocol          
% channel %
rx_wave = txrx(tx_wave1);                 
% receiver %
rx_bs = waveform2bitseq(rx_wave,SPB); % waveform=>bit 
rx_msg = bitseq2text(rx_bs);          % bit=>text 

     
\end{lstlisting}

To ensure continuity, this code is similar to that given in Task 1. The main difference is that the line tx$\_$bs = text2bitseq(tx$\_$msg) has been replaced by lower-level code that is intended to implement the function ``text2bitseq.m''.  However, there is a mistake in this code.  The students' task is to find and correct this mistake.

\emph{Step 1: Run the code.} Similar to Task 1, students click on the ``Run'' button, which returns the plots shown in Fig. \ref{t2_tx_bs} and Fig. \ref{t2_rx_wave}. By looking at the figures, students should be able to figure out that due to the error in the code, only one byte of the message is transmitted.

\emph{Step 2: Correct the code implementing text2bitseq.m.} In this step, students correct the code implementing text2bitseq.m. We first give a description of the code, and how it is supposed to convert a text string to a bit sequence. By combining this information with their previous observations of the output of the code, students must correct the code provided.  In this case, the students should replace the line ``tx$\_$bs = [byte]'' with ``tx$\_$bs = [tx$\_$bs byte]''. If they make this correction and run the code, they should see the waveforms they observed previously (Figs. \ref{s1_tx_bs} and \ref{s1_rx_wave}), confirming that they have fixed the error.

Although we do assume that students have a general understanding of procedural programming, we do not assume familiarity with MATLAB in particular.  Thus, this exercise also serves to introduce students to the vector representations and basic manipulations provided by MATLAB (e.g. creating vectors by concatenation).  In our explanations of the code, we also provide pointers to MATLAB tutorials from MathWorks, which are hosted on the site. As the course progresses, we assume that students acqurie increasing levels of familiarity with MATLAB.  The difficulty and complexity of the exercises increases accordingly.

\subsubsection{Later Tasks}

Tasks 3 and 4 are similar to Task 2. Student are asked to implement the``bitseq2waveform.m'' and ``bitseq2tex.m'' blocks of Fig. \ref{Lab_1_1}. Note that for all tasks, we make sure the initial code is syntax error free and executable. In particular, when students press the ``Run'' button, the code executes without generating MATLAB error messages.  This is extremely important because students can be easily confused by syntax errors and logic errors. This makes it more difficult for students to identify logical mistakes in the code, and would interfere with our instructional objectives.

Because it occurs early in the course, LAB1 does not contain a final evaluation task.  The first evaluation task occurs in the second week of the course, after the students have had a chance to become more familiar with the online environment for lab exercises.  In that week, the final task we ask the students to perform is an evaluation of the bit error rate as the bit rate increases.  In this task, students use the functions which they have implemented to simulate a digital communication system, and generate a plot that shows how the bit error rate increases as the bit rate increases, illustrating a fundamental trade-off in communication systems design.

\section{Design of Assessment Code and Feedback}

The assessment code (grader) determines whether students have successfully corrected or implemented MATLAB code they are required to.  If not, the code provides error messages to students to aid them in identifying the problems with their submissions.

In order to determine correctness, the grader examines the values of a relevant subset of variables in the MATLAB workspace after the student submitted code executes, and compares them with reference values that would be generated by the correct code.  Because there are different ways to implement the same algorithm, we evaluate only the results of running the code, rather than the code itself.  This is because in most cases, our primary objective is to use the process of writing the code as a means for students to better understand the algorithms and concepts being presented by the course, rather than teaching good programming style.  Nonetheless, we do seek to provide examples of good programming style in the solutions to the lab exercises, which are released on-line after the due date. The topic of efficient implementation of some of the solutions to the lab exercises has also come up in the discussion forum.

To assist students' learning progress, the grader should not only identify correct solutions, but also provide feedback to help students find their mistakes. According to the Cognitivist Framework \cite{Sperry}, there are many different misconceptions \cite{Hestenes} that can lead to incorrect solutions. In developing feedback to guide students to reach the correct solution, we paid particular attention to two categories of mistakes: errors due to misunderstanding of the communications concepts and logical errors in the code. We were not as concerned about syntax errors, as these are easily detected by the MATLAB server.

To help students develop problem solving skills, when the results of the code generated by the students do not match the expected results, the grader generates messages that provide more information than simply whether the results are correct or incorrect.  For example, for the task in which students are required to design a decoder converting a received bit sequence to a text message, the feedback provided by grader includes a display of the decoded text message, the desired text message, and information about the mismatch, e.g. mismatches in length and/or mismatches in content. These are intended to help students find errors in the code.  For example, mismatches in length might suggest problems in estimating the length of the expected text message, the for loop, or in the generation of the decoded message. We tried to provide factual information, rather than trying to guess about the source of the error, since confusing or misleading feedback messages could distract students from finding the correct solution independently.

\begin{figure}
    \begin{center}
    \scalebox{0.5}{\includegraphics{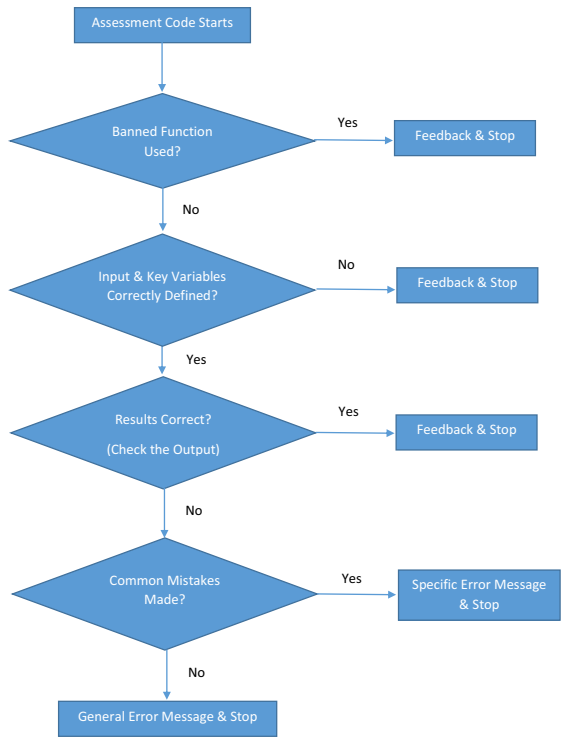}}
    \end{center}
    \caption{Flow chart for the assessment code.}
    \label{grader}
\end{figure}

The systematic assessment approach is summarized in Fig.\ref{grader} with the following steps:
\begin{itemize}
\item check for banned functions;
\item check the input and key variables to identify possible mistakes;
\item generate the correct solution and check the results (output);
\item check for common mistakes with specific error messages;
\item generate a general error message for other errors.
\end{itemize}
In the following, we use the assessment code for LAB1-Task2 to show how grader and feedback message is designed.

\subsubsection{Check for Banned Functions}
\label{SecBanned}
As described in Section \ref{SecIllustrativeExample}, students are first provided with an overview, where different components of the communication system are compartmentalized as functions. In later tasks, they will implement some of these functions using low-level MATLAB code.  To prevent students from simply calling the higher level functions provided in the overview to implement the functions in MATLAB, we defined a list of banned functions. The grader checks the students submitted code to determine whether any of these functions were used, and if so generates an error.

\subsubsection{Check the inputs and the parameters}
\label{SecInput}
As illustrated in the code provided in \ref{SecIllustrativeExample} (e.g. for LAB1 Task 2), the MATLAB scripts provided typically set up an initial set of variables that are used as inputs to or parameters for the later parts of the code to be modified by the students. Although comments provided in the code instruct students not to modify these variables, in their process of debugging or modifying the code to solve the task, students may intentionally or unintentionally change these variables.

We felt that this was undesirable for several reasons.  First, it may cause even correctly written code to malfunction. Second, in some cases, this may dramatically simplify the task to be performed by the students. Third, it complicates the design of the grader by introducing more variability into the generated output.

To avoid this, the grader checks the values of the inputs and parameters. The pseudo-code for doing this is shown in Algorithm \ref{alg:grader:1}.
\begin{algorithm}
\caption{Grader: Check parameters and input variables}
\label{alg:grader:1}
\begin{algorithmic}
\REQUIRE list of variables to check : $varNames$
\STATE
\COMMENT {Check the existence of the input/output variables:}
\FORALL {$var$ in $varNames$}
\IF {$var \notin $ workspace variables }
\PRINT ``Expected variable ` $var$ ' is missing.''
\RETURN \FALSE
\ENDIF
\ENDFOR
\STATE \COMMENT {Check the value of parameters and input variables:}
\STATE $retValue \gets $ \TRUE
\IF { $var1 \neq requiredValue1$}
\PRINT ``The variable \textit{var1} should be \textit{requiredValue1}. Do not change it.''
\STATE $retValue \gets $ \FALSE
\ENDIF
\STATE $\ldots$
\IF { $varN \neq requiredValueN$}
\PRINT ``The variable \textit{varN} should be \textit{requiredValueN}. Do not change it.''
\STATE $retValue \gets $ \FALSE
\ENDIF
\RETURN $retValue$
\end{algorithmic}
\end{algorithm}

\subsubsection{Generate the Correct Solution and Check Students' Submission}

In order to check the submitted code, we compared the results generated by the code with their expected values.  The results generated by the code were the values of variables in the MATLAB workspace after the entire script executed.  This means that if we wish to check the values of intermediate values, e.g. those generated during the execution of a for-loop, we must create variables to store these so that they are available to the grader for checking.  In some cases, we made this storage explicit to the students by including the definition and assignment of these storage values in the provided code.  We did this in cases where we felt that the students would find access to these variables beneficial in examining and debugging the results of their code.  In other cases, we hid this storage through the use of functions which defined global variables.  These global variables were given unusual names to avoid potential conflict with student generated variables.  We did this in cases where we felt that these additional storage variables would introduce unnecessary complexity to the code that would interfere with students completing the task.

In most cases, the grader generates the expected values by running a correct version of the code to be generated by the students, rather than hard coding the expected values. There are several reasons why this is necessary and/or desirable.  First, the input waveforms to the communication system are sometimes generated randomly, leading to outputs that vary randomly every time the code is run.  Second, the number of outputs that need to be checked can be prohibitively large if enumerated explicitly.  For example, sometimes the code needs to check all values of an entire waveform or simulations with a large number of intermediate variables. Third, generating the solutions on the fly by the grader enables us to make changes in the task presented to the students without the need to modify the grader after every task.

The grader must be carefully designed to avoid runtime errors. We typically checked results generated by the students' code using the following sequence:
\begin{itemize}
\item Check for the existence of a variable
\item Check properties of the variable, for example, the length if the variable is a vector.
\item Check that the value of the student generated variable matches that of the expected.
\end{itemize}
In the first two parts of this MOOC, these steps were implemented separately by code designed by instructors. In Part III, these steps were encapsulated into a function provided by MathWorks, which performs the checks on existence and size automatically as part of the assessment.

Before comparing the student and grader generated variables, it is important that the grader checks for the existence of the student generated variable.  This is because students may neglect to generate a variable in writing their code.  In this case, errors generated by comparing an existing variable generated by the grader with a non-existing variable expected to be generated by the student will generate error signals that are confusing for the students.  If the grader finds any variables that do not exist, it feeds back an error message indicating that the variable does not exist, and does not proceed with any further checks on that variable.

Even when student and grader generated variables both exist, care must be taken in making the comparison between them.

In some cases, it is sufficient to check for equality.  This is usually true when the variables to be checked can assume only a small discrete number of values (e.g. bit sequences or text messages).

However, in many cases, checking for equality will lead to false error detection.  This is often true when the variables being checked are in theory continuous valued (although in practice discrete as in any numerical simulation performed by a digital computer). For example, values generated by different correct implementations of the code may vary slightly due to differences in logically equivalent orderings of operations or in the use of equivalent implementations of the same mathematical algorithm but with different low-level operators.  In addition, the value may differ if the code includes pseudo-random number generates.  This is often encountered in this course, since modelling noise is an important part of communication systems design and analysis. Since the student code and grader code are run separately, they may generate different values.  This might be avoided by setting the seed of the random number generators to be the same, but this would cause the same ``noise'' to be generated by each simulation, which we felt would be confusing to students, so we sought to avoid this as much as possible.

In cases where checking for equality was problematic, we used thresholds on the maximum absolute or mean squared differences between the student and grader generated variables.  In the case where differences were not due to the use of pseudo-random number generators, we typically checked whether the maximum absolute element-wise difference between the variables was less than a multiple (e.g. 100) of the machine epsilon.  In cases where the differences were primarily due to the use of pseudo-random number generators, we typically used a threshold on the mean squared error, which depended upon the size of the pseudo-random numbers and their effect on the generated output.

In some cases, we also checked for what we expected to be common mistakes that might be made by students, so that we could provide more specific feedback.  For example, in Task 2 of LAB 1, students needed to implement code that converting an ASCII text message to a sequence of 8 bit binary numbers.  We anticipated that some students might mistakenly generate the sequence in reverse order.  Therefore, we also compared the student generated answer to a bit sequence in reverse order, and generated a more specific error message in this case.

If we were unable to identify the error made by the students, the grader code would simply output a generic error message indicating which of the student generated variables was not equal to its expected value.

Because we made extensive use of figures to present the data generated by the MATLAB code, in many cases the grader needed to extract the data from the figures in order to check whether the figures generated by the students were correct.  Typically, the grader first checked to make sure that the number of curves within each figure was equal to the expected number of curves, and if so, then went on to check the number of points in each curve, and then the actual values in each curve.  If any of these checks was not met, the grader generated an appropriate error signal.  This helped students to identify the mistakes in their code.

At the time we were offering the course, the MATLAB platform on edX only supported problems involving MATLAB scripts.  However, in some cases, we were interested in having students implement a function and in checking whether the code worked for all (or at least a large number of) possible inputs to the function.  In these cases, we gave student initial code consisting of a for loop, within which there was space for students to place the code of the required function.  The for loop was used to run the student code multiple times: once for each input of interest.  For the purposes of checking, we typically recorded the inputs used and the outputs generated in MATLAB cell arrays.  The grader then used the recorded inputs to generate the expected outputs, and compared the student generated outputs with these expected outputs.  This was not ideal, as the for loops added an additional layer of complexity to the code presented to the students, which was irrelevant to their understanding of what they were to implement.  In more recent versions of the MathWorks automated assessment component, there is a function type of problem available, which we intend to experiment with to see if this can reduce the complexity of the problems presented to the students.

Another situation that presented some difficulty in grading was the case where students were asked to write code that was used within a long term simulation.  For example, in one lab from Part 3 concerned with the transport layer, we asked students to implement the operation performed by the sender using the stop-and-wait protocol to transmit a number of packets in a discrete time simulation of a network that included random delays and packet loss.  The sender is given a list of packets to send.  After sending the first packet, it waits for the acknowledgement (ACK) from the receiver.  If the acknowledgement is received within a time out period, the sender then sends the next packet.  If the time out period expires before an acknowledgement is received, it resends the first packet. The sender then goes back to waiting for the corresponding acknowledgement.  This process continues until all packets are sent and acknowledged successfully.

If the students write the code of the sender correctly, then the receiver receives all of the packets in the senders list despite possible packet loss.  In order to evaluate the students' code, we created cell arrays containing all of the packets sent and all of the acknowledgements received by the sender written by the students. Based on these cell arrays, the grader checked whether the packets sent by the student's implementation of the stop and wait sender exhibited the desired behavior, and if not generated a feedback message. In particular, the grader checked for the following errors:
\begin{itemize}
\item An ACK for the current was received, but the sender did not send the next packet;
\item The timeout expired, but the sender did not resend the current packet;
\item The sender sent a packet, but no packet should have been sent.
\end{itemize}
In each case, students were given clear feedback as to which error occurred.  This helped them to identify the errors in their code. The simulation lasted long enough that most, if not all,  possible error cases would be generated.

In some cases, we also added quiz questions about the results of the MATLAB experiments, which asked students to interpret the results they observed. For example, in the lab exercises dealing with the statistical properties of the noise introduced by the communication channel, we asked the students questions about the relationship between the shape of the noise histogram and the number of samples used to generate it.  In other experiments, we asked students questions about the effect of changing the transmission distance, the bit comparison threshold or the sampling time on bit error rate.  These questions encouraged students to think critically about the results they observed, and reinforced the idea that computer simulation can be used as a tool for building understanding.

\section{Impact of MATLAB Problem Design Tools}

The quality of student experiences in taking MOOCs rely heavily upon the tools used to present material contained on the course web servers to the web browsers running on the students' computers.  Similarly, the ease of course design for MOOC instructors depends heavily upon the design tools used to develop these course materials.  This situation is more complex for courses involving computer simulation, since multiple web servers are involved.  For example, for the first two parts of the course, the design, layout and development of the course materials, quizzes, and lab exercises was done directly on the edX server.  However, for the last part of the course, we used a Learning Tools Inter-operability (LTI) platform developed by the Mathworks, where the content for the lab exercises was developed and maintained on a server maintained by Mathworks, and displayed to the students via a window embedded within the edX site.  We describe and compare these two options in more detail below.

\subsection{Development using a Custom edX/Mathworks Integration}

\begin{figure}
    \begin{center}
    \scalebox{0.4}{\includegraphics{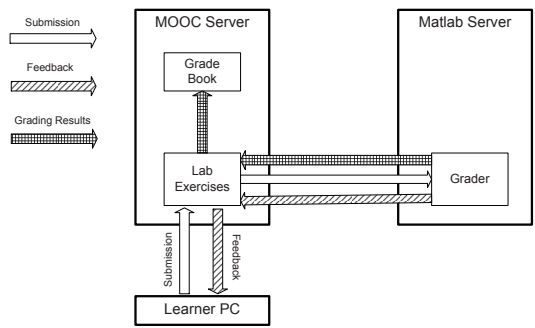}}
    \end{center}
    \caption{Network structure and data flow for lab exercises.}
    \label{Lab}
\end{figure}

For the first two parts of this course (HKUSTx: ELEC1200.1x and HKUSTx: 1200.2x), we developed the online lab exercises using a customized edX interface developed jointly by edX and Mathworks, whose network structure and data flow is shown  in Fig. \ref{Lab}.  The edX server serves as the hub for all communication between student's PC and the MATLAB server.  The initial code to be displayed to the students is encoded in an html file.  After editing the code, students click on a ``Run'' button, which sends their code to the MATLAB server for execution.  The MATLAB server returns all output generated by the code back to the edX server, which formats it for display to the students.  Once the students are satisified with their code, they submit it for assessment by clicking on a ``Check'' button.  This action sends not only the students' code, but also the grader code to the MATLAB server for execution.  If the grader code executes without detecting any errors, then the edX server is notified to update the student's record that the assignment has been completed successfully.

The advantage of this design is that all records and course materials are maintained by the edX server.  The MATLAB server simply runs the user and grader code provided to it by the edX server. This provides course developers with a single point of contact, and faciliates the integration of the lab exercises with the course flow.  For example, the edX server can maintain a record of how many times the student has submitted their code for testing and/or evaluation.  It can also update students' records immediately after they complete an exercise.

On the other hand, there are several disadvantages to this design.  Since the MATLAB server is simply executing the code passed to it by the edX server with little other auxiliary information, the MATLAB environment (e.g. available library functions) used by the server to execute the student code must be identical for all exercises.  This is dangerous, as it introduces co-dependencies between different laboratory exercises which are difficult to keep track of and maintain.  Second, because the initial and grader code are stored in html files, the original MATLAB code must be modified slightly so that it is html compliant.  The required changes are relatively minor, but it is an inconvenience, especially when developing and debugging the code.  In practice, we handled this by doing development and debugging of the lab exercises and grader using the desktop version of MATLAB, and only copying the initial and grader code into the html file on the edX server afterwards.  This two step process required additional work and checking to ensure that the final version running jointly on the edX/Mathworks servers functioned correctly.  Third, because the window in which the students write the code is maintained by the edX server, the coding environment does not feel much like the desktop version of MATLAB.

\subsection{Development using a Mathworks LTI Tool}

\begin{figure}
    \begin{center}
    \scalebox{0.4}{\includegraphics{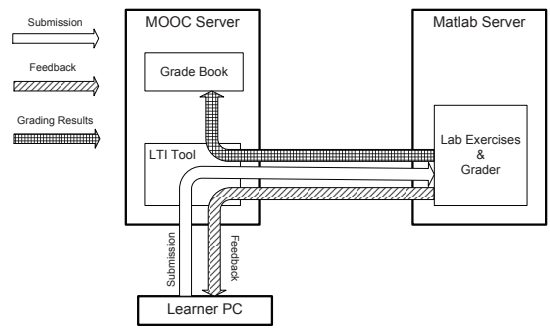}}
    \end{center}
    \caption{Network structure and data flow for lab exercises with LTI tool.}
    \label{Lab_LTI}
\end{figure}

To further facilitate the course design process, MathWorks designed a MATLAB problem creation and assessment tool using Learning Tools Inter-operability (LTI) specifications \cite{LTI}. The MOOCs platform is the LTI consumer and serves as a transparent interface between users and the MATLAB server as shown in Fig. \ref{Lab_LTI}.  In this framework, the lab instructions, the code window, and the assessment code can all be created, debugged and maintained directly on the MATLAB server.  When completing the lab tasks, students are essentially interacting directly with the MATLAB server within a window on the webpage opened by the edX server.

There are numerous advantages to this framework.  First, since the problem environment is maintained on the MATLAB server, each problem can have different MATLAB environments. This avoids the co-dependencies mentioned earlier, by providing better compartmentalization in the course design.  Second, the initial, final and grader code can be written directly in MATLAB format, avoiding the intermediate step of translating the code into HTML format.  This facilitated design and debugging of exercises directly on the online platform. However, due to delays encountered in waiting for the execution on and response from the MATLAB server, we still found initial development of the lab exercises to be easier and faster on the desktop environment.  Third, the MATLAB tool allowed for several different tests to be run on the student code.  This facilitated grader design by separating the code for different checks, rather than collating all of the checks into a single file.  It also made it clearer to the students exactly what was being checked, and in which checks the errors were being detected.  Finally, because the students were developing code within windows from the MATLAB server, the look and feel of the windows (e.g. color highlighting) was more similar to the desktop MATLAB environment.

However, there are some disadvantages to this framework.  First, the overall course content is now split across two different platforms.  Second, the integration of the student records with the online exercises is not as tight. In practice, this meant that we were unable to keep track of how many times the students had submitted or evaluated their code.  In addition, it meant that the students were not able to get credit for their work until after the submission deadline.

\section{Results}

\begin{figure}
    \begin{center}
    \scalebox{0.4}{\includegraphics{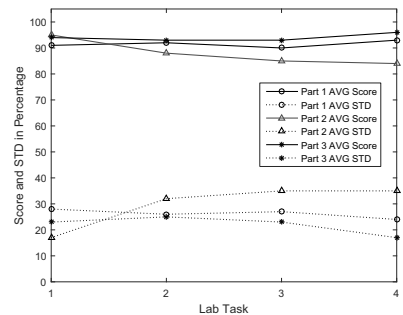}}
    \end{center}
    \caption{Comparison of students' performance in three parts by lab tasks.}
    \label{ScoreByTask}
\end{figure}

To check the effectiveness of the MATLAB Exercises, we show in Fig. \ref{ScoreByTask}, students' lab performance among four lab tasks in three parts of this course. Note that we separated the comparison by lab tasks because they are designed to address different  ILOs. We have two observations from the comparison. First, students' performance in the lab exercises confirms that the ILOs defined in Section \ref{SectionDesign}, including concept/models understanding, system design skills, problem solving skills and simulation/programming skills, were well achieved.  Second, we note that students' performance was better in Part III, than in the previous two parts.  We speculate that this is in part due to the easier and more intuitive interface provided by the LTI tools. However, we cannot rule out other confounding factors.  For example, the student population might be more skilled because of the prior practice they received in the previous sections, and the attrition of students who were performing poorly.  Second, the topics covered in the different parts varied.  A better evaluation would be based on a comparison of the same lab exercises in the two environments.  We are currently re-running Parts I and II of this MOOC using the LTI based tool, and should have this data available shortly.

To check the effectiveness of the MATLAB exercises, we also designed related survey questions. For each week, we explicitly specified the intended learning outcomes so that students know what they are expected to learn. We examine here the student responses to two statements:

\begin{itemize}
\item ``2. I learned what I expected to learn in this course.''
\item ``9. The weekly lab exercises were designed in a way that helped me learn.''
\end{itemize}

\begin{figure}
    \begin{center}
    \scalebox{0.5}{\includegraphics{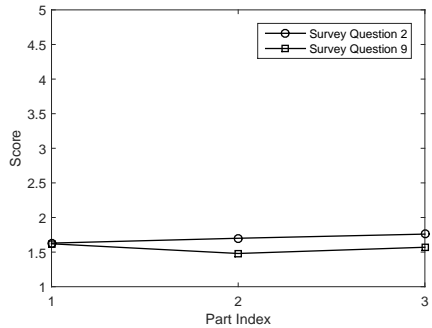}}
    \end{center}
    \caption{Results for survey questions 2 and 9.}
    \label{Survey}
\end{figure}

Students were asked to indicate the extent to which they agreed with each statements on a scale from 1 to 5, where 1 indicates strong agreement and 5 represents strong disagreement.  The first question assessed students' opinion regarding the overall effectiveness of this course. The second question was specific to the effectiveness of the lab exercises.   Fig. \ref{Survey} shows the results of this survey for three parts of this course. Most students (strongly) agree with both statements.

We also asked students to provide written comments comparing the LTI tool and the custom interface.  A representative subset of their responses is given below.
\begin{enumerate}
\item ``The new component is a lot better than the previous one: it is nice and easy to use and the error messages allow you to locate the errors much more easily.''
\item ``The feedback was helpful in steering me in corrective directions.''
\item ``I thought the MATLAB integration is great, but with the new interface it is even better, especially the ability to stretch the coding areas and write on a bigger window.''
\item ``It was better than it was in parts 1 and 2. I spent a lot of time trying out different commands to see what effects they produced.''
\end{enumerate}

It can be observed that the design tools play an important role in successful offering of online courses.

\section{Conclusions}
In this paper, we described the systematic methodology we developed for the design of MATLAB based computer simulation exercises. We described the general design philosophy, and illustrated these through a detailed example.  Key features of our approach included a structure subdivision of the lab exercise into different tasks serving different goals, the use of runtime/syntax-error free initial code, the use of figures/plots for illustration purposes, and the introduction of increasing depth and complexity in the programming tasks as the course progressed.  We also introduced our approach to the design of assessment code and feedback error messages to the students, which helped guide students to reach the correct solution and strengthen their understanding of course concepts. The achievement of the ILOs was validated by empirical measureents of students' performance.  We also described and assessed the relative advantages and disadvantages of different platforms for integrating course content and commputer simulation capabilities.  We focused in particular on a MOOC with MATLAB simulation, but anticipate that the general approach outlined here will be applicable to other online/offline teaching platforms and other simulation/programming environments.

\section*{Acknowledgement}
The authors would like to express their great appreciation to June Chan from the Center for Education Innovation, and K.S. Chiu and Almen Kwok from the Publishing Technology Center at the Hong Kong University of Science and Technology for their kind assistance in preparing the course.

\bibliographystyle{IEEEtran}
\bibliography{mooc}

\end{document}